\documentclass[twocolumn]{article}
\usepackage{times}
\usepackage{amsmath,amsthm,amssymb,verbatim,color,array,eurosym,multirow,blkarray,lscape,bm,comment,graphicx}
\usepackage[bf,small]{caption}
\usepackage[top=2.5cm,left=2.8cm,right=2.8cm,bottom=3.4cm]{geometry}
\usepackage{cancel}
\usepackage[dvipsnames]{xcolor}
\usepackage{siunitx}

\definecolor{darkgreen}{rgb}{0.0, 0.26, 0.15}
\providecommand{\keywords}[1]{\textbf{\textit{Keywords:}} #1}
\definecolor{G1}{rgb}{0.0, 0.5, 0.0}

\allowdisplaybreaks

\title{\bf Square Root Laws in Structured Fisheries}
\author{
J. Filar$^{1,*}$,
S. Streipert$^1$,
\\
$^1$ Centre for Applications in Natural Resource Mathematics\\
School of Mathematics and Physics\\
 University of Queensland  \\
j.filar@uq.edu.au, s.streipert@uq.edu.au
}

\begin{document}

\maketitle

\begin{abstract}
We introduce the term net-proliferation in the context of fisheries and establish relations between the proliferation and net-proliferation that are economically and sustainably favored. The resulting square root laws are analytically derived for species following the Beverton--Holt recurrence but, we show, can also serve as reference points for other models. The practical relevance of these analytically derived square root laws is tested on the the Barramundi fishery in the Southern Gulf of Carpentaria, Australia. A Beverton--Holt model, including stochasticity to account for model uncertainty, is fitted to a time series of catch and abundance index for this fishery. Simulations show, that despite the stochasticity, the population levels remain sustainable under the square root law. The application, with its inherited model uncertainty, sparks a risk sensitivity analysis regarding the probability of populations falling below an unsustainable threshold. Characterization of such sensitivity helps in the understanding of both dangers of overfishing and potential remedies.
\end{abstract}

\section{Introduction}
The notion of a natural ``proliferation rate'' of cells is well established in biological sciences. We extend it to fisheries and introduce the {\emph net-proliferation rate} in the context of harvested species.  Since harvest results in a reduction of the proliferation of the targeted species, this leads to the fundamental question of  characterizing the reduction level that is both  sustainable and economically viable. The contribution of this study is to demonstrate the importance of the square root function in addressing this key question in  fisheries that can be adequately described by the Beverton--Holt model \cite{BH}. We show that the following ``square root law'' relationship holds 
\begin{center}
\emph{proliferation rate \quad $\times$ \quad optimal survival rate \quad $=$ \\
\quad  square root of the proliferation rate,} 
\end{center}

\noindent
when the proliferation rate and the carrying capacity are constant. Here, optimality refers to steady state yield maximization, resulting in the maximum sustainable yield (MSY). 
%\st{In this work, we focus on fisheries that can be adequately described by Beverton--Holt type models.} 
Although this law is analytically derived for the Beverton--Holt model, we show that it can also inform  other population models. Despite the simplicity of surplus models of this type, they are critical for data-limited stock assessments \cite{Dichmont2016,Punt2015} and meta-analysis of global fisheries \cite{JABBA, Froese2016, Rosenberg2017,Worm2009}.

%resulting in the maximum sustainable yield. This suggests that a harvest rate that reduces the underlying proliferation rate (greater than 1) to the value of its square root is simultaneously sustainable and attractive for fishers. 

%The above relationship generalizes naturally to the periodic case. Furthermore, even in the stochastic case where proliferation is regarded as a random variable, its square root is essential in the analysis of the threshold risk problem. 

To formally introduce the proliferation rate in fisheries, we first note that the biomass can be normalized with respect to the species' carrying capacity $K$. Henceforth, we denote the normalized biomass, at time $t$, by $z_t$. The underlying discrete population growth model will be $z_{t+1}=G(z_t)$, for time points $t=0,1,2,\ldots$, where $G$ is a concave  smooth function with $G(0)=0$. 
%We define the {\em proliferation rate} \Sr{to be the multiplicative increase for small population levels.
We want the \emph{proliferation rate} to capture the multiplicative increase when the normalized biomass is small. Hence, mathematically, it is defined to be $\rho=G'(0)$ because the latter implies that  $z_{t+1}\approx \rho z_t$, when $z_t$ is small. The proliferation rate, so defined, satisfies $\rho=r+1,$ where $r$ denotes the {\em intrinsic growth rate} \cite{FARKAS200117,common2005ecological,waltman1983competition}.

Thus, in the generic situations of interest, the proliferation parameter $\rho >1$ is thought to capture a key biological characteristic of the species, with values close to $1$ indicating a slowly proliferating species and values much larger than $1$ corresponding to fast proliferation.  However, if the population under consideration were that of a commercially harvested fish species, the proliferation parameter $\rho$ does not capture the effect of human harvest on the species abundance. While, there are different ways of incorporating harvest into the preceding growth model, we will claim that including a multiplicative {\em (harvest) survival rate} $\gamma \in [0,1]$ is both conceptually and mathematically elegant. This is because it results in the naturally extended parametric family of models
\begin{equation}\label{eqMH}
z_{t+1} =G_\gamma(z_t)=\gamma G(z_t), \;\;\; t=0,1,\ldots \;.
\end{equation}
Thus, the unharvested model is merely the case of $\gamma=1$, namely, $100\%$ survival ($0-$harvest case).  In the case of effective harvest ($\gamma <1$), we let the {\em net-proliferation rate} be the proliferation rate in the harvested model, denoted by $\rho_{\gamma}$. Then $\rho_\gamma=G_{\gamma}'(0)=\gamma G'(0)=\gamma \rho$, which conserves the multiplicative impact of the harvest. It also shows that, effectively, human harvest reduces the numerical proliferation of the species. 

Curiously, $\sqrt{\rho}$ plays a crucial role in answering the fundamental question concerning sustainable harvest in both deterministic and stochastic settings. Starting with the deterministic setting, we derive three square-root laws linking proliferation and optimal survival rate for the Beverton--Holt model. 	

To address the question of applicability, we fit data from the Australian Barramundi population to a Beverton--Holt model incorporating random deviations.  We show that under the square root law the population remains sustainable. However, in the presence of such stochasticity, there always exists a positive probability that the population falls below a threshold. 

More generally, we define the {\em threshold-risk} as the probability that the equilibrium biomass falls below a threshold and investigate its properties in a  stochastic extension of the deterministic model. We exhibit a characteristic parametric sensitivity of this risk which, once more, depends strongly on the square root of the proliferation rate. Thus the deterministic and stochastic analyses have commonalities.

\section{The square-root laws}
First, we derive the square root laws from the deterministic Beverton--Holt model with multiplicative harvest, which is obtained by assuming that a fraction $\gamma_t \in [0,1]$ of the population survives the harvest in the time interval  $(t,t+1)$. That is, we consider 
\begin{equation}\label{eq1}
x_{t+1} = \gamma_t \frac{x_t K_t \rho_t}{K_t +(\rho_t-1)x_t }, \quad \quad \quad x_{t_0}=x_0,
\end{equation}
where $K_t>0$ is the carrying capacity at time $t$ and $\rho_t \in (1,\infty)$ is the proliferation rate.

The model described in \eqref{eq1} can also be understood as the Beverton--Holt model with harvest at the end of the period, namely
\begin{equation}\label{eq1star}
x_{t+1} =  G(x_t)- (1-\gamma_t) G(x_t),
\end{equation}
where the growth function (without harvest) is
\begin{equation}\label{Eq_Growth1}
G(x_t)=\frac{x_t K_t \rho_t}{K_t +(\rho_t-1)x_t }.
\end{equation}

An equivalent form of \eqref{eq1} has been recently studied in \cite{BoSt} which analyzed the recursion  
\begin{equation}\label{Eq_BoSt}
(1+h_t)x_{t+1}=G(x_t),
\end{equation}
where $G$ follows \eqref{Eq_Growth1}. 
We shall show that the closed-form solution of \eqref{Eq_BoSt} and the corresponding maximum sustainable yield, derived in \cite{BoSt}, acquire biological interpretation when expressed in terms of proliferation and survival rates.  

If we let $K_t>0$, $\rho_t>1$, and $\gamma_t \in [0,1]$, all be $T$-periodic, then there exists a unique, globally asymptotically stable, $T$-periodic solution to \eqref{eq1}. This follows from results in \cite{BoSt} but, for completeness, a standalone derivation is included (SI Appendix, section 2A).

When $K,\rho, \gamma$ are constant, then $T=1$ and the constant (globally attractive) equilibrium reduces to
\begin{equation}\label{optperC}
\bar{x} = \frac{\gamma \rho -1}{\rho-1} K.
\end{equation}
Despite its limitations \cite{Larkin, Mahon}, MSY is still often used to determine sustainable harvest levels in fisheries \cite{Gov1, Gov2} and will be the subject of our analysis. Thus, to obtain the optimal survival $\gamma^*$, we maximize the catch at equilibrium
$$C(\bar{x})=(1-\gamma)G(\bar{x}) = (1-\gamma)\frac{\bar{x}}{\gamma}=
\frac{(1-\gamma)}{\gamma}\frac{(\gamma \rho -1)}{(\rho-1)} K.$$
The optimal survival that results in the maximum sustainable yield is $\gamma^* = \frac{1}{\sqrt{\rho}}$ \cite{FiQiSt}. Under the optimal survival $\gamma^*$, the equilibrium in \eqref{optperC} is consistent with the optimal escapement given in Eq.\ $(59)$ of  \cite{MattHolden}.

Consequently, if the fishery were harvested so that the resulting survival rate is one over the square-root of the proliferation rate, then the catch attains the maximum sustainable yield.
This leads to our first square root law: 
\begin{center}
\emph{proliferation rate \quad $\times$ \quad optimal survival rate \quad $=$ \quad  square-root of the proliferation rate},\\[1.5mm]
\end{center}
equivalently
\begin{equation}\label{Law1}
\rho \times \frac{1}{\sqrt{\rho}}=\sqrt{\rho}.\\
\end{equation}

Although the study of MSY assumes equilibrium conditions, the same motivation can be applied to seasonally dependent population models resulting in a stable periodic steady-state. 
The authors in \cite{BoSt} addressed this by maximizing the yield with respect to $T$-periodic survival rates $\gamma_t$, under the assumption of $T$-periodic carrying capacities $K_t$ and proliferation rates $\rho_t$. 
In the special case where proliferation rates are all equal to $\rho$, the optimal $T$-periodic survival rate is 
\begin{equation}\label{optgamma}
\gamma_t^* = \frac{1}{\sqrt{\rho}}\frac{K_{t+1}}{K_t},
\end{equation}
provided that $\gamma_t^* \in (0,1)$ (SI Appendix, section 2B). Even though $K_t$ is now periodic, the optimal survival rate still depends on the square root of the proliferation rate and the ratio of the carrying capacities only at the current and successive times. This implies the second square root law for the case of periodic carrying capacities:
\begin{center}
\emph{proliferation rate \quad $\times$ \quad geometric mean of the optimal\\
\quad \quad \quad \quad \quad \quad survival rates}\\
\emph{ \quad $=$ \quad  square-root of the proliferation rate,}\\[1.5mm]
\end{center}
equivalently
\begin{equation}\label{Law2}
\rho \times \sqrt[T]{\prod \limits_{t = 0}^{T-1} \frac{1}{\sqrt{\rho}}\frac{K_{t+1}}{K_t}}=\sqrt{\rho}.
\end{equation}
Thus, in the case of a constant proliferation rate and seasonally dependent periodic carrying capacity, the product of the proliferation rate and the geometric mean of the optimal periodic survival rates, once again, yields the square-root of the proliferation rate. \\

The above can be extended to the case of seasonally dependent survival rates $\rho_t$. By adapting results in \cite{BoSt} (SI Appendix, section 2B), the optimal, $T$-periodic harvest survival rate at time $t$ that maximizes the yield is given by
\begin{equation}\label{optgamma_gen}
\gamma_t^* = \frac{1}{\sqrt{\rho_t}} \frac{K_{t+1}}{K_t}
\left[  \frac{\sqrt{\rho_{t}}+1}{\sqrt{\rho_{t+1}}+1}\right],
\end{equation}
provided that $\gamma_t^*\leq 1$ for all $0\leq t <T$. While this expression for the optimal survival rate contains more factors, the square root of the proliferation rates is, as before, crucial. Once again, only the current and successive time points are important. 

This leads to the third square root law (SI Appendix, section 2C):
\begin{center}
\emph{geometric mean of proliferation rates\\
\quad $\times$ \quad geometric mean of optimal survival rates \\
\quad $=$ \quad square-root of the geometric mean of proliferation rates.}\\
\end{center}
Even though the analytical results are specifically derived for the Beverton--Holt model, the definition of a net-proliferation rate is more general, allowing tests of the square-root laws for other models. In the case of the Pella--Tomlinson model \cite{Pella_Tomlinson}, the (unharvested) growth function $G$ for the normalized population is given by 
$$G(z_t) = z_t + \frac{r}{m-1}z_t\left( 1 -z_t^{m-1}\right),$$
where $m>1$ and $0 \leq r \leq 1$. This includes the Fox model \cite{Fox} when the shape parameter $m$ tends to $1$ from above and the discrete logistic model when $m = 2$. 
The proliferation rate of this model is $\rho = G'(0) = 1+\frac{r}{m-1}$. In the Pella--Tomlinson model, the survival rate that maximizes the catch at equilibrium is  $\gamma_{msy} = \frac{m}{\rho(m-1)+1}$. 

\begin{figure}[h!]
\includegraphics[scale=0.16]{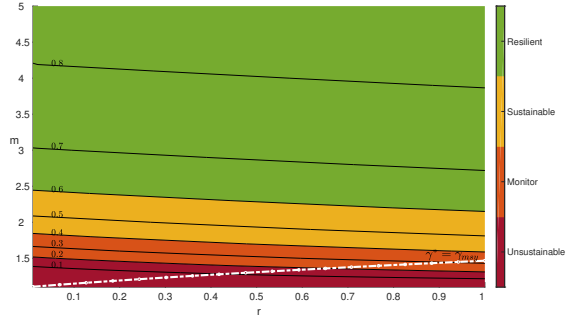}
\caption{Contour plot of population levels at equilibrium for the Pella--Tomlinson model dependent on parameters $r$ and $m$ with the survival following the square root law. Color-coding is consistent with the sustainability characterization provided by the Queensland government \cite{Gov1}. For $m$ greater than $1.5$, the resulting biomass levels can reach a biomass level of $0.4$, while for $m>2$, even low growth rate values $r$ yield sustainable biomass levels. Parameter values above the white curve imply higher survival rates for $\gamma^*$ than for $\gamma_{msy}$.}\label{Compmodel}
\end{figure}

Not surprisingly, $\gamma_{msy}$ differs from $\gamma^* = \frac{1}{\sqrt{\rho}}$. However, applying the first square root law as a rule-of-thumb will lead to sustainable biomass levels for a wide range of parameter values, as illustrated in Figure \ref{Compmodel}. Applying the square root law to the positive equilibrium of the Pella--Tomlinson model yields the equilibrium biomass $z^* = \sqrt[m-1]{\frac{\sqrt{\rho}}{\sqrt{\rho}+1}}$. 

From Figure \ref{Compmodel}, it is apparent that for $m>1.5$, the biomass levels fall, dependent on $r$, in the sustainable or at least monitoring category based on classification proposed in \cite{Gov1}.
That implies that some species of Gadiformes $(m=1.729)$ and Scorpaeniformes ($m=1.970$), see  \cite{Thorson}, are examples of taxonomic groups falling within sustainable levels under the square root law. 
However, for parameter values $m$ close to 1, the resulting biomass levels under the square root law are characterized as  ``unsustainable''. Even when population levels are within the red regions, for parameter pairs above the white curve $\gamma^*=\gamma_{msy}$, the square root law $\gamma^*$ is in fact more conservative than the model's suggested survival rate $\gamma_{msy}$.
This suggests that for a wide range of  parameters, the square root law can be used as a conservative rule-of-thumb.

\section{Demonstration using Barramundi data}
\begin{figure}[h!]
    \centering
    \includegraphics[width=7cm,height=3cm]{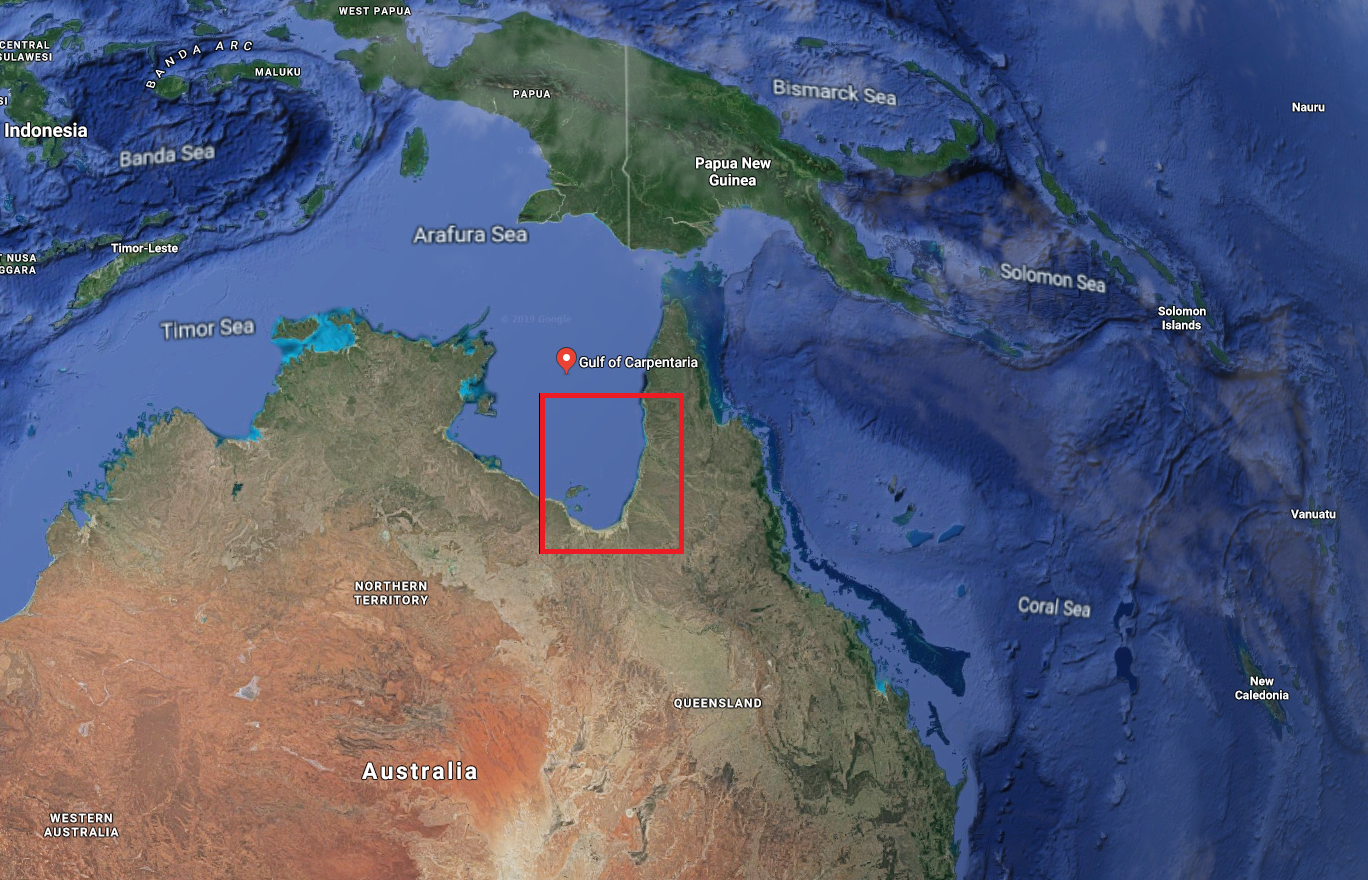}
    \caption{Satellite image of the Southern Gulf of Carpentaria in Australia \cite{GoogleEarth}. The highlighted region extends from \ang{13} South near the Watson River on Western Cape York to the Queensland/Northern Territory border at $\sim$ \ang{138} East, see \cite{StockAs_Barra2019}.}
    \label{SGulf_map}
\end{figure}

The theoretical results can also be linked to data-driven models. As a demonstration, we consider the population of Lates calcarifer (Barramundi/Asian sea bass)  and show that if the survival rate satisfies the first square root law, the species remains viable. We focus on the Southern Gulf of Carpentaria Barramundi  fishery, see Figure \ref{SGulf_map}, which is  economically valuable with annual catch of around 800 tons. 

\begin{figure}[h!]
    \centering
    \includegraphics[scale=0.25]{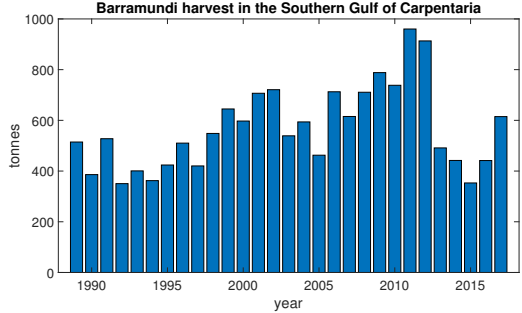}
    \caption{Annual catch from 1989 to 2017 in tonnes provided by the Queensland Department of Agriculture and Fisheries in 2018 for the Southern Gulf of Carpentaria (SGulf), Australia. These raw data were obtained from the compulsory CFISH logbook data \cite{StockAs_Barra2019}.}
    \label{Catch1}
\end{figure}

For that region, we used commercial catch data from a compulsory logbook (CFISH) displayed in Figure  \ref{Catch1}. We also used an abundance index time series from 1989 to 2017, recently calculated in \cite{StockAs_Barra2019}. 

For this application of the Beverton--Holt model in \eqref{eq1}, we assumed that parameters are time invariant, namely, $K=K_t, \; \rho=\rho_t$ and $\gamma=\gamma_t,$ at all times. To account for random  deviations from the underlying deterministic recurrence, we multiplied \eqref{eq1} by $e^{\eta_t},$ where $\eta_t \sim N(0,\sigma^2)$. 

\begin{figure}[h!]
    \centering
    \includegraphics[scale=0.22]{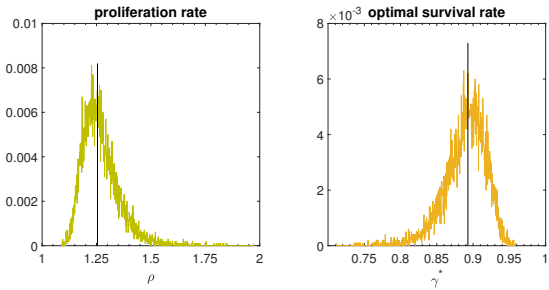}
    \caption{Posterior distribution of the parameter $\rho$ obtained from the MCMC fit of the Beverton--Holt model to the Barramundi abundance index of the Southern Gulf of Carpentaria. The median value of this posterior distribution, highlighted by the vertical line, is 1.2545. The posterior distribution of the optimal survival rate $\gamma^* = \frac{1}{\sqrt{\rho}}$ is given in the right panel, with a median value of 0.8928.}
    \label{FigBarra2}
\end{figure}

The model fitting was done according to a Bayesian paradigm  \cite{beamish2009future,pitcher2012reinventing, council1998improving, PUNT1997} based upon \cite{JABBA} (SI Appendix, Section 1).  During the Markov Chain Monte Carlo (MCMC) procedure we saved every 5$^{th}$ iteration of model parameters, generating a sample of $4$-tuples $\psi=(K,\rho,\sigma^2, \varphi)$. Here, $\varphi$ denotes the normalized population level at the beginning of the time series.  
The left panel of Figure \ref{FigBarra2} displays the resulting posterior distribution of the proliferation rate $\rho$ from these saved $4$-tuples and the right panel plots the corresponding $\gamma^*$ values, calculated according to the first square root law. The black vertical lines indicate the respective median values $\rho_m \approx 1.25$ and $\gamma^*_m\approx 0.89$. For each saved $4$-tuple $\psi$, the index of agreement introduced by Willmot \cite{Willmott1984, Krause}, was calculated as an indicator of the goodness-of-fit. The distribution of these indicator values suggested acceptable fits with its $5^{th}$ and $95^{th}$ percentile values of 0.6244 and 0.8686 and a median value of $ 0.7763$.

\begin{figure}[h!]
\centering
\includegraphics[scale=0.4]{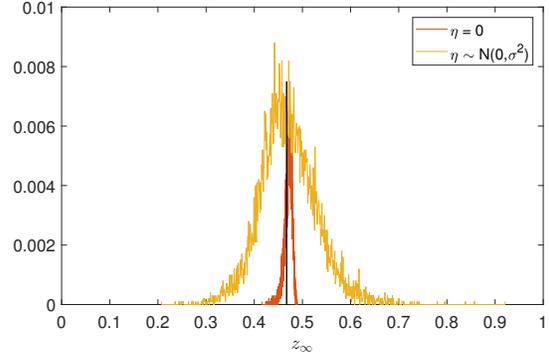}
\caption{The yellow curve represents the population level at equilibrium $z_{\infty}$ when applying the square root law to the model \eqref{eqSQRJAB} with the posterior distributions for $\rho, K, \eta, \varphi$ obtained from the MCMC. The vertical line is at the median value of approximately 0.467.
The red curve displays the simulated  equilibrium assuming a posterior distribution of $\eta$ equal to $0$ (to remove the effects of model deviations) and, consequently, has smaller variance. Comparing the two curves, the distribution considering model deviations is slightly right skewed, while the distribution of equilibria without model deviations is moderately left-skewed.}\label{FigBarra}
\end{figure}

To simulate the effect of the square root law, for each parameter combination $\psi$, we calculated the survival rate $\gamma^*=\frac{1}{\sqrt{\rho}}$ and applied the recurrence
\begin{equation}\label{eqSQRJAB}
z_{t+1} = \gamma^* \frac{\rho z_t}{1+ (\rho - 1)z_t}e^{\eta_t}, \quad \quad z_{t_0} = \varphi,
\end{equation}
for the biomass divided by $K$. Since the population $z_t$ stabilized after 100 iterations, $z_{2090}$ was assumed to represent the limiting population denoted by  $z_{\infty}$. The resulting distribution of these values is displayed in Figure \ref{FigBarra} (yellow curve). The red distribution in the same figure corresponds to the noiseless case $\eta_t\equiv 0.$ The medians of the two curves agree closely with the analytical equilibrium under the square-root law, namely, $\frac{\rho_m \gamma^*_m -1}{\rho_m-1}=\frac{\sqrt{\rho_m} -1}{\rho_m-1}\approx 0.47$ (see \eqref{optperC}, with $K=1$).

Thus, we see that when fitting the data to \eqref{eqSQRJAB} and applying the first square root law, the (normalized) biomass reaches sustainable levels above 0.4 \cite{Gov1} for most parameter configurations in both the noisy and noiseless cases.  Given that the parameters in $\psi$ can be viewed as realizations of random variables with associated (posterior) distributions, there is typically a positive probability that the population falls below a sustainable threshold.
For instance, we see from the yellow curve in Figure \ref{FigBarra}, that $z_{\infty}$ could fall below 0.25, even though that is unlikely.
This ``risk'' of falling below a threshold if parameters are chosen from a probability distribution is explored analytically in the next section.

\section{Square Root Law in Risk Analysis}

Henceforth, we assume that the proliferation rate $\rho$ is a continuous random variable with support on a subset of $(1,\infty)$ and $F_\rho$ is its cumulative distribution function (cdf).
We aim to determine the risk that the population in steady-state is falling below a given positive threshold $\delta$.   Extending a preliminary investigation in  \cite{FiQiSt}, we assume that the carrying capacity at time $t$, $K_t$, and the harvest survival rate at time $t$, $\gamma_t$, are both $T$-periodic. 
We require that the support of $\rho$ is such that the survival rate lies between $0$ and $1$.

\subsection{Risk analysis under optimal harvest}

First, let us assume that the harvest rate is maximizing the sustainable yield. Under the optimal harvest survival rate given in \eqref{optgamma}, the $T$-periodic steady-state solution can be simplified to (SI Appendix, section 2D)
\begin{equation}\label{optpersol}
\bar{x}_t^* = \frac{K_t}{\sqrt{\rho}+1}.
\end{equation}
Hence, the threshold-risk for the $T$-periodic solution at time $t$ under optimal harvest is given by the period independent quantity
\begin{equation}\label{Thold}
P( \bar{x}_t^* <\delta) =
1-F_{\rho}\left(\frac{(K_t-\delta)^2}{\delta^2}\right)
\end{equation}
if $\delta < K_{t}$. If the threshold $\delta \geq K_t$, then the risk is one. 

\begin{figure}[h!]
    \includegraphics[scale=0.25]{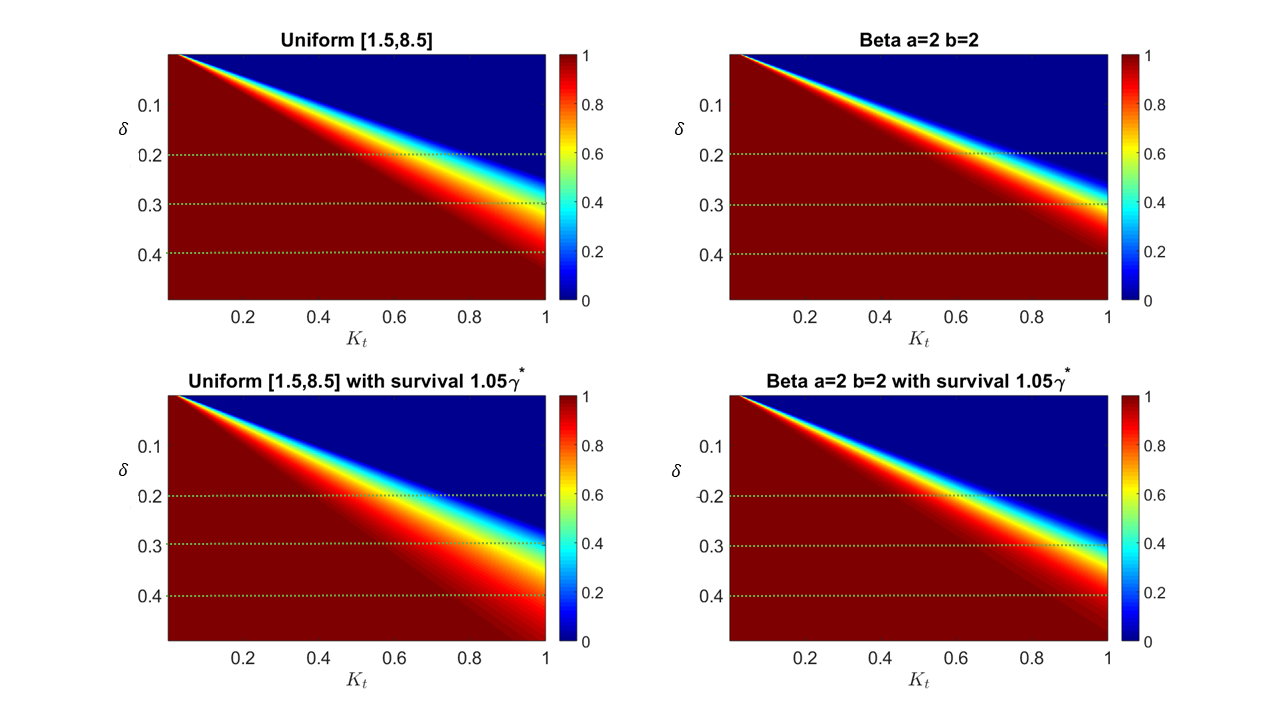}
    \caption{Behaviour of $P(\bar{x}^*_t<\delta)$ for  $\rho$ uniformly distributed on $[1.5,8.5]$ for the left panels. The right panels assume that the random variable $\frac{\rho-1.5}{7}$ follows a Beta distribution with parameters $\alpha=2=\beta$ so that  the support of $\rho$ is again $[1.5,8.5]$. The risk is color coded, ranging from dark blue to crimson as risk increases. The top row calculates the risk corresponding to the optimal survival $\gamma^*$. The bottom row illustrates the risk if the survival rate were increased by $5\%$. It  shows that the wedges (areas of intermediate risk) are now wider, replacing levels of higher risk. 
    Also the low risk (dark blue) areas are larger and straight lines of constant color risk have more negative slopes and reach lower values of the threshold $\delta$.}
    \label{fig:1}
\end{figure}

In Figure \ref{fig:1}, we demonstrate a characteristic sensitivity of the threshold-risk \eqref{Thold} to parameters, $\delta, K_t$. Wedge-like regions are observed, characterizing the relative ease (or difficulty) of changes between high and low threshold-risk for two  distributions of the random variable $\rho$ with support $[1.5,8.5]$. The thinness of the wedges in Figure \ref{fig:1} in certain regions of the parameter space conveys two messages. On the one hand, it underscores the dangers of overfishing even when maximum sustainable harvest (from the deterministic model) is used. On the other hand, it suggests relatively easy remedies. For instance, Figure \ref{fig:1} shows that just a small increase in survival rate to $1.05\gamma^*$ considerably thickens these wedges and generally reduces regions of high risk.

\subsection{Risk analysis for constant greedy harvest}

Suppose that, instead of using the optimal harvest $\gamma^*$, a greedy deviation $\gamma=\theta \gamma^*$, where $\theta \in \left(0,1\right)$ is implemented.
%Setting $\gamma = \theta \frac{1}{\sqrt{\rho}}\frac{K_{t+1}}{K_t}$, 
The corresponding periodic solution for $t=0,1, \ldots T-1$ is (SI Appendix, section 2E)
\begin{equation}\label{xtheta}
\bar{x}_t^{\theta} = K_t \frac{ \sqrt{\rho} \, \theta-1}{\rho-1}.
\end{equation}

The threshold-risk for the $T$-periodic solution under a constant greedy deviation from the optimal harvest can again be expressed only in terms of the cdf $F_\rho$. It can be shown (SI Appendix, 2F) that the risk is either $1$ or  
\begin{equation*}
P( \bar{x}_t^{\theta}<\delta)
    = 
H(F_{\rho})+ F_{\rho}\left(\frac{1}{\theta^2}\right).  
\end{equation*}
The significance of this expression is that, as in \eqref{Thold}, the cdf of a non-linear transformation of the random variable $\rho$ reduces to an explicit expression in the (known) cdf of $\rho$. 
The threshold-risk is still memoryless because $H(F_{\rho})$ only depends on parameters at the current time $t$ (SI Appendix, 2F).

\subsection{Risk analysis for periodic greedy harvest}

When the greedy deviation parameter becomes time dependent, then the memoryless property is lost. Consider the survival at time $t$ to be $\gamma_t = \theta_t \gamma_t^*$, where $\gamma_t^*$ is the optimal survival at time $t$ and $\theta_t \in (0,1)$. 
Let 
$\vec{\theta}=(\theta_0, \theta_1, \ldots, \theta_{T-1})$ and $\bar{\theta}$ be the product of its entries.
Simplifying the solution (SI Appendix, section 2G),
 we observe that the $T$-periodic solution $\bar{x}_t^{\vec{\theta}}$ is again linear in $K_t$ but depends on time dependent greed $\vec{\theta}$.

Even though the risk of the solution falling below a threshold $\delta$ is now more complex, we can link it to the roots of a higher order polynomial in the square root of the proliferation rate. To be precise, we have (SI Appendix, section 2H) 
\begin{multline}\label{Greedprob}
P(\bar{x}_t^{\vec{\theta}}<\delta) = 
P(\{h(\sqrt{\rho}>0\} \, \cap \, \{\sqrt{\rho}^T \bar{\theta}>1 \}) \\
+ P(\sqrt{\rho}^T \bar{\theta}\leq 1), 
\end{multline}
where 
\begin{equation}\label{Poly}
h(\sqrt{\rho})=\sum_{i=0}^{T+1} a_i\sqrt{\rho}^i >0
\end{equation}
is a polynomial of order $T+1$ in $\sqrt{\rho}$.  The coefficients of $h$ depend on the parameter values in the $T-$period cycle and  hence the risk is not memoryless.

\begin{figure}[h!]
    \centering
    \includegraphics[scale=0.4]{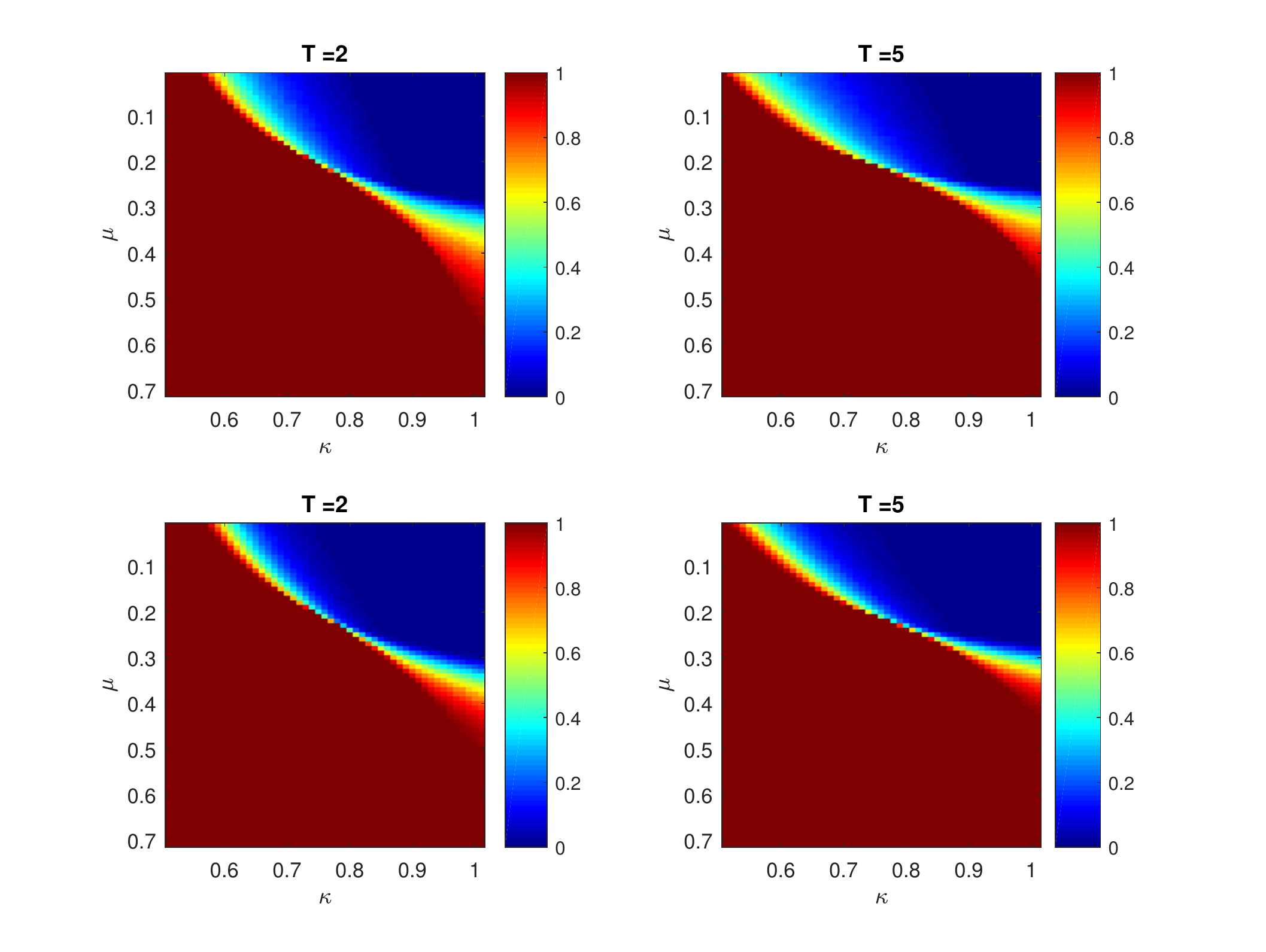}
    \caption{Behaviour of the probability $P(\bar{x}^{\vec{\theta}}_1<\delta)$ that the $T$-periodic solution under time dependent greed falls below the threshold $\delta$.  This risk depends on $\mu = \frac{\delta}{K_1}$ and the $T$-periodic greed. Here, the relation $\theta_t = \kappa + (1-\kappa)\cos\left( \frac{2\pi}{T}t  \right)$ was assumed. Using a sampling method to estimate this threshold-risk. The top row assumes a uniformly distributed $\rho$ on $[1.5,8.5]$ and distinguishes between a $2$-periodic greed and $5$-periodic greed. The bottom row, assumed that $\frac{\rho-1.5}{7}$ follows a beta distributed with parameters $\alpha=2=\beta$ so that the support of $\rho$ is again $[1.5,8.5]$. For both choices of the period and both distributions, the risk exhibits a curved 2-wedge with intermediate risk spreading on either side of a highly sensitive risk location.}
    \label{periodgreed}
\end{figure}

The threshold-risk can now be calculated as the probability of the inequality \eqref{Poly}. In Figure \ref{periodgreed} we, once again, demonstrate the sensitivity of this risk to small parameter changes. We note that the wedges observed earlier are now split and curved. 
We conjecture that the above splitting is a consequence of the fact that the inequality in \eqref{Poly} could be satisfied in multiple parts of the support of $\rho$. Fortunately, we show that there can be at most three such parts, irrespective of the value of $T$.    

Evidently, the threshold-risk assessment is linked to finding the roots of the polynomial $h$, whose characteristics are influenced by the coefficients.  The sign of most coefficients can be easily determined. While $a_1<0$, $a_i>0$ for $2\leq i \leq T-1$ and $a_{T+1}>0$, the sign of $a_0$ and $a_T$ could attain both, positive and negative values (SI Appendix, section 2I).
Further, it is shown that if $a_0\leq 0$, the population falls below the threshold $\delta$ at time $t$ and the risk is one.  For the case $a_0>0$ we exploit the classical Descartes' rule of signs \cite{Descarteproof, Descarteproofsimple} to obtain the risk of one if $a_T\geq 0$ (SI Appendix, section 2J).

Hence the threshold-risk lies between $0$ and $1$ only when $a_0>0$ and $a_T<0$.
In such a case,  utilizing Descartes' rule once more, leads to the conclusion that $h$ can have only $0$, $2$ or $4$ positive roots.   
Consequently, there can be at most three intervals where $h(\sqrt{\rho})$ is positive. Note that for a general $(T+1)$-order polynomial that number could be as high as  $\lceil \frac{T+2}{2} \rceil$.  Thus the computational effort in calculating threshold-risk is not prohibitive even when $T$ is large.

\section{Conclusion}
We present three square root laws for the newly introduced net-proliferation rate, associated to the maximum sustainable yield for a family of species whose population dynamics can be adequately captured by the Beverton--Holt type models. However, we also demonstrate that when the square root law is applied to populations captured by the Pella--Tomlinson model this still results in sustainable population levels, for a wide range of parameter values. 
Although surplus models of this type can be criticized for their simplicity, they are critical in  data-limited stock assessments \cite{Dichmont2016,Punt2015} and meta-analysis of global fisheries \cite{JABBA, Froese2016, Rosenberg2017,Worm2009}. 
The Beverton-Holt recurrence possesses  many mathematically desirable properties  which are preserved when a multiplicative harvest survival rate is incorporated. Analysis of this harvested Beverton--Holt model exposes, hitherto unknown, significance of the square root of the underlying proliferation rate of the species.   

It is established that  the net-proliferation rate -- introduced as the proliferation rate of the harvested model -- is intimately linked to the proliferation rate of the species via three square root laws. 
The most general of these, in the periodic case, states that:  

\begin{center}
\emph{geometric mean of proliferation rates \quad $\times$ \quad geometric mean of optimal survival rates \\
\quad $=$ \quad square root of the geometric mean of proliferation rates.}\\
\end{center}
\vspace{-0.3cm}
 In the words of A.\ Hastings \cite{Hasting} these laws can, perhaps, be viewed as ``exact solutions to approximate (simple) models rather than approximate solutions to more detailed models''. This claim is addressed further by applying the non-periodic version of the above law to an Australian Barramundi fishery. Despite the complexity of the species \cite{Wang1, Balston} and the inherent randomness in data driven studies, the application of the square root law simulated sustainable biomass levels. In the process, a  risk of the population falling below sustainable levels was observed. This led to a deeper analysis of the sensitivity of such threshold-risk.

More precisely, we considered the proliferation rate to be a random variable. Then, we were able to reduce the risk of the abundance falling below a threshold to the probability of a polynomial in the square root of the proliferation rate being positive. High parameter sensitivity is to be expected, guided by the location of at most four positive roots of that polynomial.

\textbf{Acknowledgement} We thank the Queensland Department of Agriculture and Fisheries for providing Barramundi data, and Australian Research Council Grant DP180101602. We are also indebted to our colleagues Dr.\ M.\ Holden, Dr.\ W-H.\ Yang and Dr.\ J.\ Robins for many valuable discussions and insights.

\bibliography{arxiv_version} 
\bibliographystyle{siam}

\newpage

\onecolumn

\begin{small}

\section*{Appendix}
\subsection*{A1: Application to Lates Calcarifer}
Given two data inputs, observed catch from 1989 to 2017 (Catch\textunderscore SGulf.csv) and abundance index in form of standardised catch (CPUE\textunderscore SGulf.csv) for the same time frame \cite{StockAs_Barra2019}, we fit the Beverton--Holt type model %\st{The base of the estimation process was the JABBA software described in} \cite{JABBA}, \st{available at: \emph{https://github.com/JABBAmodel}}. \st{However, the Pella-Tomlinson recursion used in} \cite{JABBA} \st{has been replaced by Eq.\ $(2)$ in the main text file.  The model is scaled by considering the density with respect to the normalized biomass, $z_t=\frac{x_t}{K}$, where $K$ is the carrying capacity and $x_t$ denotes the population's density at time $t$. 
%In the starting year 1989, the biomass $x$ was set equal to $\varphi K$,  a parameter to be estimated within the model} \st{The stochastic form of the model differs from the model in} \cite{JABBA} \st{and is given by}
\begin{equation*}
    z_t = \begin{cases}
    \varphi e^{\eta_t}& t=1989\\
    \left(\frac{z_{t-1}}{(1-r)+rz_{t-1}} - \frac{C_{t-1}}{K}\right)e^{\eta_t}& 1989<t<2018
    \end{cases},
\end{equation*}
where $r=\frac{(\rho-1)}{\rho}$ and $z_t$ is the normalized biomass at time $t$ with respect to the carrying capacity $K$. The observed catch at time $t$ is denoted by $C_t$. The additional factor $e^{\eta_t}$ captures the process error, that is assumed to be log-normally distributed with mean zero and variance $\sigma_{\eta}^2$. To account for measurement errors, the observations $I_t$ of the biomass index, fitted to the observed abundance index, are assumed to be noisy with log-normal distribution of mean zero and variance $\sigma^2_{\xi}$.

The JAGS ('Just Another Gibbs Sampler') R-software (\emph{http://mcmc-jags.sourceforge.net/}), following the model set up in \cite{JABBA}, was used with the joint probability distribution over the parameters $\psi:=\{K, r,\sigma^2_{\eta}, \sigma^2_{\xi}, q, \varphi\},$ the process errors ${\bf \eta}=(\eta_{1989}, \eta_{1990}, \ldots, \eta_{2017})$ and observation errors ${\bf I}=(I_{1989}, I_{1990},\ldots, I_{2017} )$. By Bayes' Theorem, the joint posterior distribution over all unknown parameters, given the data, is  
\begin{multline*}
p(\psi, {\bf \eta}\, | \,{\bf I}) \sim p(K)p(r)p(\varphi)p(\sigma^2_{\eta})p(q)p(\sigma^2_{\xi})p\left( z_{1989}\, \mid \, \varphi, K, \sigma^2\right)p\left(I_{1989} \, \mid \, z_{1989}, q, \eta_{1989}, \sigma^2_{\xi_{1989}} \right) \times \\
\times \left[ \prod_{k=1990}^{2017} p\left(z_{k} \, \mid \, z_{k-1}, K, r, \varphi, \sigma^2_{\eta} \right)p\left(I_{k} \, \mid \, z_{k}, q, \eta_k, \sigma^2_{\xi_{k}} \right)\right],
\end{multline*}
where 
\begin{align*}
    K \sim & \,\mbox{lognorm}(200000,1) \\ %\quad \quad \quad (\mbox{lognormal with mean and coefficient of variation}) \\
    r \sim & \, \mbox{lognorm}(0.42, 0.37)\\
    \varphi \sim & \,\mbox{lognorm}(1,0.25) \\
    q \sim & \,\mbox{Uniform}(0,100) \\
    \sigma^2_{\eta} \sim & \,\frac{1}{\mbox{gamma}(4,0.01)} \\
    \sigma^2_{\xi} \sim &\, \frac{1}{\mbox{gamma}(0.001,0.001)}+0.04 \\
\end{align*}
The choice of the prior of the process error is due to adequate estimation performance \cite{Ono2012, Thorson2014} and corresponds to a mean of 0.059 and cv at 0.28 \cite{MeyerMiller1999, MeyerMiller2000}.

Using two Markov Chain Monte Carlo chains with 30000 iterations each, saving every 5th parameter combination and using a burn-in of 5000 steps. The saved MCMC parameter combinations have been used to produce the histogram of the posterior distribution of $\rho$ in Figure \ref{FigBarra2} of the main text. These saved parameter combinations also served the simulation of $z_{\infty}$, illustrated in Figure \ref{FigBarra}. For each parameter combination, the optimal survival rate $\gamma$ was calculated by simply setting $\gamma^* = \frac{1}{\sqrt{\rho}}$ and running the model 
\begin{equation*}
    z_t = \begin{cases}
    \varphi e^{\eta_t}& t=1989\\
    \gamma^* \frac{z_{t-1}}{(1-r)+rz_{t-1}} e^{\eta_t}& 1989<t<2090
    \end{cases}.
\end{equation*}
Checking for convergence, $z_{2090}$ satisfied the convergence condition of $z_{2091}-z_{2090}<0.000001$ and was hence assumed to be the converging (normalized) biomass under the square root law, denoted by $z_{\infty}$. The resulting histogram for $z_{\infty}$ for the saved MCMC parameter combinations is plotted in Figure \ref{FigBarra} in the main text.

\subsection*{A2: Analytical Derivations}

\subsection*{A Solution to the harvested Beverton--Holt model}
%The $T$-periodic solution to the harvested Beverton--Holt model is given by Equation (28) in \cite{BoSt}, which is presented as 
To obtain the solution to the harvested Beverton--Holt equation for $K, \rho, \gamma$ being $T$-periodic, we note that for $x_t>0$  
$$
 x_{t+1} = \frac{\gamma_t K_t \rho_t x_t }{K_t + (\rho_t-1)x_t} \quad \quad \leftrightarrow \quad \quad \frac{1}{x_{t+1}} = \frac{1}{\gamma_t \rho_t}\frac{1}{x_t} + \frac{\rho_t-1}{\gamma_t \rho_t K_t} \quad \quad \leftrightarrow \quad y_{t+1} = \frac{1}{\gamma_t \rho_t}y_t + \frac{\rho_t-1}{\gamma_t \rho_t K_t}
$$
with $y_t = \frac{1}{x_t}$. To simplify the notation, we define $a_t:=\rho_t \gamma_t$ (which is also $T$-periodic) and $b_t:=\frac{\rho_t-1}{K_t}$. Then the recurrence in $y_t$ has the solution 
$$y_t = y_0\prod_{k=0}^{t-1} a_k^{-1} + \sum_{k=0}^{t-1} \frac{b_k}{a_k}\prod_{j=k+1}^{t-1} a_j^{-1}$$
for the initial condition $y_0=\frac{1}{x_0}$. To find the $T$-periodic solution, we set $\bar{y}_{t}=\bar{y}_{t+T}$ to obtain
\begin{equation*}
\bar{y}_{t+T} = y_0\left(\prod_{k=0}^{t-1} a_k^{-1}\right) \left(\prod_{k=t}^{t+T-1} a_k^{-1} \right)+ \sum_{k=0}^{t-1} \frac{b_k}{a_k}\left(\prod_{j=k+1}^{t-1} a_j^{-1}\right) \left(\prod_{j=t}^{t+T-1} a_j^{-1} \right)+
\sum_{k=t}^{t+T-1} \frac{b_k}{a_k}\left(\prod_{j=k+1}^{t+T-1} a_j^{-1}\right).
\end{equation*}
Using additionally the periodicity of $a_k$, $\prod_{k=t}^{t+T-1}a_k = \prod_{k=0}^{T-1} a_k$ which can then be factored out from the first two terms to obtain
\begin{align*}
\bar{y}_{t+T}&= \left(\prod_{k=0}^{T-1} a_k^{-1}\right)  \underbrace{\left[y_0\left(\prod_{k=0}^{t-1} a_k^{-1}\right)+ \sum_{k=0}^{t-1} \frac{b_k}{a_k}\left(\prod_{j=k+1}^{t-1} a_j^{-1}\right)\right]}_{=\bar{y}_t} +
\sum_{k=t}^{t+T-1} \frac{b_k}{a_k}\left(\prod_{j=k+1}^{t+T-1} a_j^{-1}\right).
\end{align*}

Setting $\bar{y}_{t+T}$ equal to $\bar{y}_t$ and solving for $\bar{y}_t$ yields the $T$-periodic solution $\bar{y}_t$ as 
\begin{equation*}%\label{Eq}
\bar{y}_t = \frac{1}{1-\prod_{k=0}^{T-1} a_k^{-1}} \sum_{k=t}^{t+T-1}  b_k \left(\prod_{j=k}^{t+T-1} a_j^{-1}\right).
\end{equation*}
Resubstituting yields
\begin{equation}\label{Solper}
\bar{x}_t = \left(1-\prod_{k=0}^{T-1} a_k^{-1}\right) \left[ \sum_{k=t}^{t+T-1}  b_k \left(\prod_{j=k}^{t+T-1} a_j^{-1}\right)\right]^{-1}
=
\left(\left(\prod_{k=0}^{T-1} a_k\right)-1\right) \left[ \sum_{k=t}^{t+T-1}  \frac{b_k}{a_k}\left(\prod_{j=t}^{k}a_j\right)\right]^{-1}
\end{equation}
%where we have used that 
%$$\sum_{k=t}^{t+T-1} b_k \prod_{j=k+1}^{t+T-1}a_j^{-1} =\left[\prod_{j=t}^{t+T-1}a_j^{-1}\right]\left[\sum_{k=t}^{t+T-1} b_k \prod_{i=t}^{k} a_i    \right]  $$
Conversely, one can show that this $\bar{x}_t$ solves the harvested Beverton--Holt model and is $T$-periodic. This is consistent with the solution obtained in \cite{BoSt}, where $\bar{x}_t$ was proved to be globally asymptotically stable.

\subsection*{B Optimal Harvest}
Following the same approach as Theorem 3.10 in \cite{BoSt}, we maximize the catch over one period evaluated at the optimal periodic solution by obtaining an upper bound that is achieved only for the proposed optimal harvest. 
Using results in A, the first step is to realize that the catch over one period is of the form 
\begin{align}\label{Harv1}
\sum_{t=0}^{T-1} (1-\gamma_t) G(\bar{x}_t)&= \sum_{t=0}^{T-1} (1-\gamma_t) \frac{\bar{x}_{t+1}}{\gamma_t} = \left(\prod_{k=0}^{T-1} \rho_k \gamma_k -1\right) \sum_{t=0}^{T-1} \frac{(1-\gamma_t)}{\gamma_t} \frac{1}{ \sum_{j={t+1}}^{t+T} \left[ \prod_{k=t+1}^{j} \rho_k \gamma_k  \right]\frac{\rho_j-1}{\rho_j K_j \gamma_j} }\notag \\
&=\left(\prod_{k=0}^{T-1} \rho_k \gamma_k -1\right) \sum_{t=0}^{T-1} \frac{(1-\gamma_t)}{\gamma_t} \frac{1}{ \sum_{j={t+1}}^{t+T} w_{j,t}  x_{j,t} }
\end{align}
with 
$x_{j,t} = \left[ \prod_{k=t+1}^{j} \sqrt{\rho}_k \gamma_k  \right]\frac{(\sqrt{\rho}_j+1)}{\sqrt{\rho}_j K_j \gamma_j}$
and 
$w_{j,t} = \frac{\sqrt{\rho_j}-1}{\sqrt{\rho_j}}\prod_{\tau = t+1}^j \sqrt{\rho}_\tau$. 
We apply Jensen's inequality to the convex function $f(x)=x^{-1}$ to obtain
\begin{equation}\label{Harv2}
\frac{1}{\sum_{j=t+1}^{t+T} \frac{w_{j,t}}{W_t}x_{j,t}} \leq \sum_{j=t+1}^{t+T} \frac{w_{j,t}}{W_t}\frac{1}{x_{j,t}} \quad \quad \leftrightarrow 
\quad \quad 
\frac{1}{\sum_{j=t+1}^{t+T} w_{j,t} x_{j,t}} \leq \sum_{j=t+1}^{t+T} \frac{w_{j,t}}{W_t^2}\frac{1}{x_{j,t}}
\end{equation}
where $W_t = \sum_{j=t+1}^{t+T} w_{j,t}$. Applying \eqref{Harv2} to every term in \eqref{Harv1} results in \begin{equation}\label{Harv3}
\sum_{t=0}^{T-1} (1-\gamma_t) G(\bar{x}_t)  \leq \left(\prod_{k=0}^{T-1} \rho_k \gamma_k -1\right) \sum_{t=0}^{T-1} \frac{(1-\gamma_t)}{\gamma_t}  \sum_{j={t+1}}^{t+T} \frac{w_{j,t}}{W_t^2}  \frac{1}{x_{j,t}}.
\end{equation}
It can be verified that $W_t$ is independent of $t$ due to the periodicity of $\rho_t$, in particular,
$$W_t = \prod_{j=0}^{T-1} \sqrt{\rho_j}-1.$$
Again using the periodicity of $K_t, \gamma_t$ and $\rho_t$, we get
$$\sum_{t=0}^{T-1}\frac{(1-\gamma_t)}{\gamma_t} \sum_{j=t+1}^{t+T} \frac{w_{j,t}}{x_{j,t}} = 
\left(\prod_{j=0}^{T-1} \gamma_j^{-1}-1\right)
\sum_{j=1}^{T} \frac{(\sqrt{\rho_j}-1)}{\sqrt{\rho_j}+1}K_j.$$
For the last equation, we have interchanged the summations using the rule
$$\sum_{t=0}^{T-1}\sum_{j=t+1}^{t+T} \alpha_{j,t} = \sum_{j=1}^{T}\sum_{t=0}^{j-1}\alpha_{j,t} + \sum_{j=T+1}^{2T-1} \sum_{t=j-T}^{T-1}\alpha_{j,t}$$

Using these two equations in \eqref{Harv3} results in 
$$\sum_{t=0}^{T-1} (1-\gamma_t) G(\bar{x}_t)  \leq \frac{\left(\prod_{k=0}^{T-1} \rho_k \gamma_k -1\right)}{
\left[\prod_{i=0}^{T-1}\sqrt{\rho}_{i}-1\right]^2} \left(\prod_{k=0}^{T-1} \gamma_k^{-1} -1\right) 
\sum_{j=1}^T  (\sqrt{\rho_j}-1) \frac{K_j }{(\sqrt{\rho}_j+1)}.$$
%\leq \sum_{j=1}^T  (\sqrt{\rho_j}-1) \frac{K_j }{(\sqrt{\rho}_j+1)}
One can verify that 
\begin{equation}\label{Ineq2}
\frac{\left(\prod_{k=0}^{T-1} \rho_k \gamma_k -1\right)}{
\left[\prod_{i=0}^{T-1}\sqrt{\rho}_{i}-1\right]^2}\left(\prod_{k=0}^{T-1} \gamma_k^{-1} -1\right)  \leq 1
\end{equation}
as this inequality is equivalent to 
$$\prod_{k=0}^{T-1} \gamma_k^{-1}  + \prod_{k=0}^{T-1}\gamma_k \rho_k 
-2\prod_{i=0}^{T-1}\sqrt{\rho}_i = \left(\prod_{k=0}^{T-1}\sqrt{\gamma_k}\sqrt{\rho_k} - \prod_{k=0}^{T-1}\sqrt{\gamma_k}^{-1}       \right)^2 \geq 0.$$
That implies that the catch over one period is bounded above by $\sum_{j=1}^T  (\sqrt{\rho_j}-1) \frac{K_j }{(\sqrt{\rho}_j+1)}$. To show that the proposed optimal harvest 
\begin{equation}\label{optgamma}
\gamma_t^* = \frac{1}{\sqrt{\rho}_t}\frac{K_{t+1}}{K_t}\frac{\sqrt{\rho_t}+1}{\sqrt{\rho_{t+1}}+1}  
\end{equation}
achieves that bound, it is sufficient to show that it attains equality in \eqref{Harv3} and \eqref{Ineq2}. Since it is easy to verify the latter, we focus on the equality in \eqref{Harv3}. From Jensen's inequality, it is sufficient to show that $x_{j,t}$ are independent of $j$. This is the case because
\begin{multline*}
\left[\prod_{k=t+1}^j \sqrt{\rho_k}\gamma_k^*\right] \frac{\sqrt{\rho}_j+1}{\sqrt{\rho_j}K_j \gamma_j^*}=    \left[\prod_{k=t+1}^j \sqrt{\rho_k}
    \frac{1}{\sqrt{\rho}_k}\frac{K_{k+1}}{K_k}\frac{\sqrt{\rho_k}+1}{\sqrt{\rho_{k+1}}+1}\right]
    \frac{\sqrt{\rho}_j+1}{\sqrt{\rho_j}K_j \frac{1}{\sqrt{\rho}_j}\frac{K_{j+1}}{K_j}\frac{\sqrt{\rho_j}+1}{\sqrt{\rho_{j+1}}+1}}\\
    =\left[\prod_{k=t+1}^j 
    \frac{K_{k+1}}{K_k}\frac{\sqrt{\rho_k}+1}{\sqrt{\rho_{k+1}}+1}\right]
    \frac{(\sqrt{\rho_{j+1}}+1)}{ K_{j+1}}    =
    \left[ 
    \frac{K_{j+1}}{K_{t+1}}\frac{\sqrt{\rho_{t+1}}+1}{\sqrt{\rho_{j+1}}+1}\right]
    \frac{(\sqrt{\rho_{j+1}}+1)}{ K_{j+1}}=
    \frac{\sqrt{\rho_{t+1}}+1}{K_{t+1}}.
        \end{multline*}
Thus, the upper bound is reached, making $\gamma^*$ the optimal harvest survival rate and the corresponding catch $C(\gamma^*) = \sum_{j=1}^T  (\sqrt{\rho_j}-1) \frac{K_j }{(\sqrt{\rho}_j+1)}$ the maximum sustainable yield.
Note that if $\rho_t \equiv \rho$, we recover from \eqref{optgamma} 
$\gamma_t^* = \frac{1}{\sqrt{\rho}}\frac{K_{t+1}}{K_t}$. Thus the second square root law in the main document follows easily from this simplification.

\subsection*{C Simplification of Law 3:}
For $K$ and $\rho$ being $T$-periodic, we have $K_T=K_0$ and $\rho_T=\rho_0$ and realizing that 
$$\prod_{t=0}^{T-1} \frac{K_{t+1}}{K_t}=\frac{K_T}{K_0}=1, \quad \quad \prod_{t=0}^{T-1} \frac{\sqrt{\rho_t}+1}{\sqrt{\rho_{t+1}}+1}=\frac{\sqrt{\rho_0}+1}{\sqrt{\rho_{T}}+1}=1$$
yields, using the optimal survival rate \eqref{optgamma}, the third square root law 
\begin{equation}\label{Law3}
%\begin{split}
\left(\prod_{t=0}^{T-1} \rho_t\right)^{\frac{1}{T}} \times \left(\prod_{t = 0}^{T-1} \gamma_t^*\right)^{\frac{1}{T}}
=\left(\prod_{t=0}^{T-1} \rho_t\right)^{\frac{1}{T}}\times \left(\prod_{t = 0}^{T-1} \frac{1}{\sqrt{\rho_t}}\cdot \frac{K_{t+1}}{K_t}\cdot \frac{\sqrt{\rho_{t}}+1}{\sqrt{\rho_{t+1}}+1}\right)^{\frac{1}{T}} \\
%&=\left(\prod_{t=0}^{T-1} \rho_t\right)^{\frac{1}{T}} \times \left(\left(\prod_{t = 0}^{T-1} \frac{1}{\sqrt{\rho_t}}\right)\cdot \frac{K_{T}}{K_0}\cdot   \frac{\sqrt{\rho_{0}}+1}{\sqrt{\rho_{T}}+1}\right)^{\frac{1}{T}} =\left(\prod_{t=0}^{T-1} \rho_t\right)^{\frac{1}{T}} \times \left(\left(\prod_{t = 0}^{T-1} \frac{1}{\sqrt{\rho_t}}\right)\right)^{\frac{1}{T}}
= \sqrt{\left( \prod_{t=0}^{T-1} \rho_t\right)^{\frac{1}{T}}}.
%\end{split}
\end{equation}
%using the periodicity of $\rho$ and $K$.\\

\subsection*{D Simplification of the solution under optimal harvest}
We recall that the periodic solution is given in \eqref{Solper}. Letting $\rho_t\equiv \rho$, $K_t$ be $T$-periodic, and using the optimal survival rate $\gamma_t^*$ given in \eqref{optgamma}, we have $a_t = \rho \gamma_t^* = \sqrt{\rho}\frac{K_{t+1}}{K_t}$ and $b_t = \frac{\rho-1}{K_t}$. We note that 
$$\prod_{j=t}^k a_j = \sqrt{\rho}^{k-t+1} \frac{K_{k+1}}{K_{t}}.$$
Hence the solution in \eqref{Solper} simplifies to
\begin{equation*}
\bar{x}_t^* 
 = \left(\sqrt{\rho}^{T}-1\right) \left[ \sum_{k=t}^{t+T-1}  \frac{\rho-1}{\sqrt{\rho} K_{k+1}}\cdot \sqrt{\rho}^{k-t+1}\frac{K_{k+1}}{K_t} \right]^{-1} 
 = \left(\sqrt{\rho}^{T}-1\right)\frac{K_t}{\rho-1}\cdot  \left[ \sum_{k=t}^{t+T-1}  \sqrt{\rho}^{k-t} \right]^{-1}=  \frac{K_t}{\sqrt{\rho}+1},
\end{equation*}
where we have used the geometric sum formula in the last step.

\subsection*{E Solution under greedy harvest conditions}\label{Optz}
Let $\rho>1$ and let $K_t$ be $T$-periodic. Suppose now that the survival rate $\gamma_t=\theta \gamma_t^*$ for $\theta \in (0,1)$, then 
$a_t = \rho \theta \gamma_t^* = \theta \sqrt{\rho}\frac{K_{t+1}}{K_t}$ and $b_t = \frac{\rho-1}{K_t}$ in the solution \eqref{Solper}. We note that 
$$\prod_{j=t}^k a_j = \theta^{k-t+1}\sqrt{\rho}^{k-t+1} \frac{K_{k+1}}{K_{t}},$$
which simplifies the solution in \eqref{Solper} to
\begin{align*}
\bar{x}_t^{\theta} 
&=\left( \theta^{T}\sqrt{\rho}^{T}-1\right) \left[ \sum_{k=t}^{t+T-1}   \theta^{k-t} \frac{1}{K_t} \sqrt{\rho}^{k-t}  (\rho-1)\right]^{-1}=\left( \theta^{T}\sqrt{\rho}^{T}-1\right) \frac{K_t}{(\rho-1)} \left[ \sum_{k=t}^{t+T-1} \theta^{k-t}\sqrt{\rho}^{k-t}  \right]^{-1}\\
&=\left( \theta^{T}\sqrt{\rho}^{T}-1\right) \frac{K_t}{(\rho-1)} \left[ \frac{1-\sqrt{\rho}\theta}{1-\left(\sqrt{\rho}\theta\right)^{T}} \right]
= K_t \frac{\sqrt{\rho}\theta -1}{(\rho-1)}.
\end{align*}

\subsection*{F Risk analysis under greedy harvest conditions}
Using the expression in Section \ref{Optz}, we see immediately that for $\sqrt{\rho}\theta\leq 1$, the risk is one, so
\begin{equation*}
    P(\bar{x}_t^{\theta}< \delta ) = 
    P\left( \left\{K_t \frac{\sqrt{\rho}\theta -1}{(\rho-1)}< \delta \right\} \cap \{\sqrt{\rho}\theta>1\}\right) +P(\sqrt{\rho}\theta\leq 1)
    =P\left( \left\{ g(\sqrt{\rho}) >0\right\} \cap \{\sqrt{\rho}\theta>1\}\right) +P(\sqrt{\rho}\theta\leq 1)
    \end{equation*}
    where $g(\sqrt{\rho}) = \mu_t \rho - \mu_t + 1 - \sqrt{\rho}\theta $ is a polynomial in $\sqrt{\rho}$ of order $2$ with $\mu_t = \frac{\delta}{K_t}$. %Let $z_1 \leq z_2$ be the roots of $g$. 
    The roots of $g$ are
$   z_1 =\frac{\theta - \sqrt{\Delta}}{2\mu_t} \leq z_2=\frac{\theta+ \sqrt{\Delta}}{2\mu_t}$ where $\Delta = \theta^2 - 4\mu_t(1-\mu_t)$. Then
\begin{equation*}
P\left( \{g(\sqrt{\rho})>0\} \, \cap \,\{\sqrt{\rho} \theta >1\}\right) =
\begin{cases}
1- F_{\rho}\left( \frac{1}{\theta^2} \right) \, & \, \Delta\leq 0\\
F_{\rho}(z_1^2) - F_{\rho}\left( \frac{1}{\theta^2}  \right)+1-F_{\rho}(z_2^2)
\, & \, \frac{1}{\theta}<z_1, \Delta>0\\
1-F_{\rho}(z_2^2) \, & \, z_1 \leq \frac{1}{\theta}\leq z_2, \Delta>0\\
F_{\rho}\left( \frac{1}{\theta^2}   \right) \, & \,  \frac{1}{\theta}> z_2, \Delta >0,\\
\end{cases}
 \end{equation*}
 where $F_{\rho}$ is the cumulative distribution function of the random variable $\rho$. Further, since $P(\sqrt{\rho}\theta \leq 1) = P\left(\rho \leq \frac{1}{\theta^2}\right)=F_{\rho}\left( \frac{1}{\theta^2}\right)$, the risk can be explicitly expressed in terms of the cdf of $\rho$.

\subsection*{G Solution under time-dependent greedy harvest}
Let $\rho$ be constant, $K$ be $T$-periodic and the survival rate $\gamma_t=\theta_t \gamma_t^*$, where $\theta_t \in (0,1)$ and $\gamma^*_t$ is given in \eqref{optgamma}, then by \eqref{Solper},
\begin{equation*}
\bar{x}_t^{\vec{\theta}} = \left\{ \left[\prod_{k=0}^{T-1}\theta_k \gamma^*_k\right]\rho^{T}-1\right\}\cdot \left[ \sum_{j=t}^{t+T-1} \left[ \prod_{k=t}^j \theta_k \gamma_k^*\rho \right]  \frac{\rho-1}{\rho \theta_j \gamma_j^* K_j}\right]^{-1}.
\end{equation*}
Substituting $\gamma^*_t = \frac{1}{\sqrt{\rho}}\frac{K_{t+1}}{K_t}$ and using the periodicity of $K_t$, we have
\begin{align}
\bar{x}_t^{\vec{\theta}}& =\left\{ \left[    \prod_{k=0}^{T-1}\theta_k \right]\sqrt{\rho}^{T}-1\right\}\cdot \left[ \sum_{j=t}^{t+T-1} \left[ \prod_{k=t}^j \theta_k \right] \frac{K_{j+1}}{K_t} \sqrt{\rho}^{j-t}  \frac{(\rho-1)}{\theta_j K_{j+1}}\right]^{-1}\notag \\
&=K_t\frac{\left[    \prod_{k=0}^{T-1}\theta_k \right]\sqrt{\rho}^{T}-1}{\rho-1}\cdot \left[ \sum_{j=t}^{t+T-1} \left[ \prod_{k=t}^j \theta_k \right]  \sqrt{\rho}^{j-t}  \frac{1}{\theta_j }\right]^{-1} =K_t\frac{\left[    \prod_{k=0}^{T-1}\theta_k \right]\sqrt{\rho}^{T}-1}{\rho-1}\cdot \left[ 1+\sum_{j=1}^{T-1} \left[ \prod_{k=0}^{j-1} \theta_{k+t} \right]  \sqrt{\rho}^{j}  \right]^{-1}. \label{Solper2}
\end{align}
This formulation reveals that in order to obtain a feasible (positive) population, $\sqrt{\rho}^T\prod_{k=0}^{T-1}\theta_k>1$. The solution $\bar{x}_t^{\vec{\theta}}$ is then linear in $K_t$ but does not have the memoryless property as before, due to the $\theta_{k}$-terms.

\subsection*{H Relation to higher order polynomial}
We note first that if $\sqrt{\rho}^T \bar{\theta}\leq 1$, where $\bar{\theta} = \prod_{k=0}^{T-1}\theta_k$, then $\bar{z}_t^{\vec{\theta}}\leq 0$ and the risk is one. To discuss the case if $\sqrt{\rho}^T \bar{\theta}>1$, we utilize the expression of the periodic solution \eqref{Solper2} derived in Section G. %Multiplying both sides of the 
For $\mu_t = \frac{\delta}{K_t}$, we obtain
\begin{align*}
\bar{x}_t^{\vec{\theta}}&=K_t \frac{\sqrt{\rho}^T \bar{\theta}-1}{\rho-1} \left[ 1 + \sum_{j=1}^{T-1} \left[ \prod_{k=0}^{j-1}\theta_{k+t} \right] \sqrt{\rho}^j\right]^{-1}<\delta \quad
 \iff  \quad  \frac{\sqrt{\rho}^T \bar{\theta}-1}{(\sqrt{\rho}^2-1)} \frac{1}{\left[ 1 + \sum_{j=1}^{T-1} \left[ \prod_{k=0}^{j-1}\theta_{k+t} \right] \sqrt{\rho}^j\right]}<\mu_t\\
\iff & \quad \quad  \sqrt{\rho}^T \bar{\theta}-1 <\mu_t (\sqrt{\rho}^2 - 1)+\mu_t(\sqrt{\rho}^2 - 1) \sum_{j=1}^{T-1} \left[ \prod_{k=0}^{j-1}\theta_{k+t} \right] \sqrt{\rho}^j  \\
\iff & \quad \quad  h(\sqrt{\rho}):=-\sqrt{\rho}^T \bar{\theta}+1+
\mu_t \sqrt{\rho}^2 - \mu_t+\mu_t   \sum_{j=1}^{T-1} \left[ \prod_{k=0}^{j-1}\theta_{k+t} \right] \sqrt{\rho}^{j+2}- \mu_t  \sum_{j=1}^{T-1} \left[ \prod_{k=0}^{j-1}\theta_{k+t} \right] \sqrt{\rho}^{j}>0.  
\end{align*}

Setting $\omega_{j-1}:= \prod_{k=0}^{j-1} \theta_{k+t}$ and combining the two summations, we obtain a more compact expression 
\begin{align*}
h(\sqrt{\rho}) &= 1 - \mu_t  +\mu_t  \sum_{j=3}^{T-1} \left(\omega_{j-3} -\omega_{j-1}\right) \sqrt{\rho}^{j}
+ \left(\mu_t \omega_{T-3} - \bar{\theta}\right)  \sqrt{\rho}^{T}
+ \mu_t\omega_{T-2} \sqrt{\rho}^{T+1}- \mu_t \omega_{0}  \sqrt{\rho} + \mu_t(1-  \omega_1) \sqrt{\rho}^2\\
&=\sum_{i=0}^{T+1} a_{i}\sqrt{\rho}^i,
\end{align*}
where
\begin{itemize}
\item The coefficient for $\sqrt{\rho}^0$ is $a_0=1-\mu_t$. Since $\mu_t$ is positive, the sign of $a_0$ could be positive or negative.
\item The coefficient for $\sqrt{\rho}^1$ is $a_1=-\mu_t \omega_0 = -\mu_t \theta_{t}$, which is negative since $\mu_t, \theta_t$ are positive.
\item If $T>2$, the coefficient for 
$\sqrt{\rho}^2$ is $a_2=\mu_t(1- \omega_1) = \mu_t\left(1-\prod_{k=0}^1 \theta_{t+k}\right)$ which is positive since $\mu_t$ is positive and $\theta_t$ is between zero and one.
\item If $T>3$, the coefficient for
$\sqrt{\rho}^i$ is $a_i=\mu_t(\omega_{j-3} -\omega_{j-1}) = \mu_t \left( \prod_{k=0}^{j-3} \theta_{k+t} - \prod_{k=0}^{j-1} \theta_{k+t}\right)=\mu_t \left( 1 -\theta_{t+j-1}\theta_{t+j-2}\right) \prod_{k=0}^{j-3} \theta_{k+t}$ for $3\leq i \leq T-1$. By the same argument as above, these coefficients are positive.
\item The coefficient for $\sqrt{\rho}^T$ is 
$a_T=\mu_t \omega_{T-3} - \bar{\theta} = \mu_t \prod_{k=0}^{T-3}\theta_{t+k} - \prod_{k=0}^{T-1}\theta_{t+k} =\left( \mu_t - \theta_{t+T-1}\theta_{t+T-2}\right)\prod_{k=0}^{T-3} \theta_{t+k}$. Since we only know that $\mu_t, \theta_t$ are positive and $\theta_t$ is between zero and one, the sign of this coefficient is not pre-determined.
\item The coefficient of the term with the highest power $\sqrt{\rho}^{T+1}$ is $a_{T+1}=\mu_t \omega_{T-2} = \mu_t \prod_{k=0}^{T-2}\theta_{t+k}$, which is positive since $\mu_t$ and $\theta_t$ are positive.
\end{itemize}

\subsection*{I Positivity of Polynomial Coefficient $a_0$}\label{Pos1}

We will show that if the coefficient $a_0\leq 0$ in $h(\sqrt{\rho})$, then the risk of the population falling below the threshold $\delta$ is one. If $a_0\leq 0$, i.e. $\mu_t\geq 1$, then $a_{T}\geq 0$, since $\theta_t \in (0,1)$ for all $t$ and hence, $a_i>0$ for $i>1$ and $a_0,a_1<0$. By Descartes' rule of signs, the function $h$ has exactly one positive real root. Further, we have $h(0)=a_0\leq 0$ and 
\begin{align*}
h(1) &= 1-\mu_t - \mu_t \theta_{t} + \mu_t - \mu_t \theta_{t}\theta_{t+1} + \mu_t\sum_{i=3}^{T-1} \left[ \prod_{k=0}^{i-3} \theta_{k+t}-\prod_{k=0}^{i-1} \theta_{k+t} \right] -\bar{\theta}+\mu_t \prod_{k=0}^{T-3} \theta_{k+t}  +\mu_t \prod_{k=0}^{T-2} \theta_{k+t}\\
& = 1 - \mu_t \left\{ \theta_{t} + \theta_{t}\theta_{t+1} + \sum_{i=3}^{T-1} \left(\prod_{k=0}^{i-1} \theta_{k+t}\right) \right\}
 + \mu_t\sum_{i=3}^{T-1} \left( \prod_{k=0}^{i-3} \theta_{k+t} \right) -\bar{\theta} + \mu_t \prod_{k=0}^{T-3} \theta_{k+t}   +\mu_t \prod_{k=0}^{T-2} \theta_{k+t}\\
& = 1 - \mu_t   \sum_{i=1}^{T-1} \left(\prod_{k=0}^{i-1} \theta_{k+t}\right)
 + \mu_t\sum_{i=3}^{T-1} \left( \prod_{k=0}^{i-3} \theta_{k+t} \right) -\bar{\theta} + \mu_t \left(\prod_{k=0}^{T-3} \theta_{k+t}\right)  +\mu_t \left(\prod_{k=0}^{T-2} \theta_{k+t}\right)\\
 &= 1 - \mu_t   \sum_{i=0}^{T-4} \left(\prod_{k=0}^{i} \theta_{k+t}\right)
 + \mu_t\sum_{i=3}^{T-1} \left( \prod_{k=0}^{i-3} \theta_{k+t} \right) -\bar{\theta} = 1 - \mu_t   \sum_{i=0}^{T-4} \left(\prod_{k=0}^{i} \theta_{k+t}\right)
 + \mu_t\sum_{i=0}^{T-4} \left( \prod_{k=0}^{i} \theta_{k+t} \right) -\bar{\theta} = 1  -\bar{\theta} >0.
\end{align*}
Therefore, the only positive root is between $[0,1)$ which implies that $h(\sqrt{\rho})>0$ for all $\rho>1$. Since the probability of falling below $\delta$ is equivalent to the probability of $h(\sqrt{\rho})>0$, as established in H, the risk is one.

\subsection*{J Negativity of Polynomial Coefficient $a_T$}
By Section \ref{Pos1}, we now focus on $a_0>0$. Assume $a_T>0$, then the only negative coefficient is therefore $a_1$ and we will show that the risk is one. 
The fact that $h(1)=1-\bar{\theta}$ implies 
\begin{equation}\label{newJ1}
\sum_{k=1}^{T+1}a_k = h(1)-a_0=1-\bar{\theta} - a_0 =1-\bar{\theta} -1+\mu_t =\mu_t - \bar{\theta}\geq \mu_t-\theta_{t+T-1}\theta_{t+T-2}=a_T>0.
\end{equation}
Setting $z=\sqrt{\rho}$ and taking the derivative of $h$ evaluated at $z=1$ yields
\begin{equation*}%\label{EqDer1}
    h'(1) = \sum_{i=1}^{T+1} ia_{i}z^{i-1}\Big|_{z=1}
= \sum_{i=1}^{T+1} ia_{i} 
\geq \sum_{i=1}^{T+1}a_{i}\stackrel{\eqref{newJ1}}{>}0.
\end{equation*}
The polynomial $h'(z)$ has only real coefficients and $a_1$ is the only negative $a_i$. Hence, by  Descartes' rule of signs, $h'(z)$ has exactly one positive real root.
If the probability of the event $\{\bar{x}^{\vec{\theta}}_t<\delta\}$ were less than one, there would have to be an interval $(\rho_1, \rho_2)$ with $1<\rho_1< \rho_2$ such that $h(z)<0$. 
But since $h(1)>0,$ $h'(1)>0$ and $h(z) \to \infty$ as $z \to \infty$, if there were an interval where $h(z)<0$ that would imply the existence of both a local maximum and a minimum of $h(z)$ to the right of $z=1$. This means that $h'(z)$ has at least two positive roots which yields a contradiction.

\end{small}

\end{document}